\documentstyle[manuscript,aps]{revtex}

\newcommand{\beq}{\begin{equation}}
\newcommand{\eeq}{\end{equation}}
\newcommand{\beqa}{\begin{eqnarray}}
\newcommand{\eeqa}{\end{eqnarray}}
\newcommand{\beqar}{\begin{eqnarray*}}
\newcommand{\eeqar}{\end{eqnarray*}}

\def \la {\langle}
\def \ra {\rangle}

\def \h {{\cal H}}

\begin{document}

\title{\bf\Large Unruh Effect with Back-Reaction - 
A First Quantized Treatment}
\author{
B. Reznik\footnote{ e-mail: reznik@t6-serv.lanl.gov} \\ 
T-6 MS B288, Los Alamos National Lab., 
Los Alamos, NM 87545}

\maketitle 

\begin{abstract}

{\small   
We present a first quantized treatment of the back-reaction on 
an accelerated particle detector. The evaluated transition amplitude
for detection agrees with previously obtained results. 
}
\end{abstract}
\newpage


\section{Introduction}

In view of the close connection of  black hole radiation
and  acceleration radiation  [1-3] it is
reasonable to expect that some of the   
difficulties regarding the former case, could be mirrored 
and examined on the latter simpler problem.
In particular, it can be anticipated that the modifications 
of the Unruh effect due  
to the detector's recoil and the quantum smearing 
may have similar consequences for the Hawking effect.

Surprisingly, since the work of Unruh \cite{unruh} two decades
ago, this problem has attracted little attention.
In his original paper \cite{unruh}, Unruh suggested a  
two-field model for a finite mass particle detector. 
Two scalar fields $\Psi_M$ and $\chi_{M'}$, 
of masses $M$ and $M'=M+\Omega$, respectively, were 
taken to represent two states of a detector.
When the coupling to a scalar field
$\phi$ is given by $\epsilon\phi\chi_{M'}\Psi_M$, 
``excitation'' of this detector 
corresponds to a detection of a $\phi$ particle of energy
$\Omega$.

The two-field model was used recently by Parentani
\cite{parentani} to study the consequences
of recoil and smearing on the Unruh effect.
In this work two charged scalar fields where accelerated
by means of a classical constant external electric field.
The transition amplitude was obtained to  
first order in the coupling. 
Parentani showed that when the recoil and smearing are taken into
account, up to $\Omega/M$ corrections,
the transition amplitude is modified only by a phase. 
Therefore, to this order the thermal distribution and the Unruh
temperature are unmodified. 
The new aspect that Parentani emphasized was 
the appearance of an additional phase in the transition
amplitude which generates decoherence in successive emissions
by the detector.

In this article we shall present an alternative first quantized treatment
and rederive Parentani's result.
We shall be studying the same problem of a finite mass particle
detector in a constant electric field, but we use 
a different toy model.
The accelerating detector is described as a first quantized,
charged, point-like particle in a constant electric field with
internal energy levels.
A quantum version of (canonically conjugate) future and
past Rindler horizon operators  can be introduced \cite{qframe},
which facilitate the calculation of the transition amplitude and
provides a simple intuitive physical picture of the 
recoil.

The model is then used to obtain  the 
transition amplitude.
In the zeroth order, neglecting  $\Omega/M$ correction, we recover 
Parentani's result of an unmodified exact thermalization
in the Unruh effect. The phase found in \cite{parentani},  that
induces decoherence effects, is also obtained. In our approach,
the phase is directly obtained by acting on the wave function
with the horizon shift operator which induces the recoil.

The article is organized as follows.
In Section 2. we introduce our model for an accelerated particle
detector and construct the Rindler horizon operators.  
In Section 3, the transition amplitude is calculated and 
the result compared with that of ref. \cite{parentani}.
The recoil and quantum smearing are explicitly manifested in
Section 4, and a qualitative simple picture of the recoil 
in terms of the shifting operators is demonstrated.
In the following
we adopt the units in which
$\hbar=k_B=c=G=1$.

\section{Accelerated Detector with  Finite Mass}

In this section we  present a model for a particle 
detector of finite mass which takes into account 
also the quantum nature of the detector's trajectory.

Consider a particle detector of rest mass $M$ and
charge $q$ 
in a constant external electric field $E_x$ in $1+1$
dimensions.
Let us describe the internal structure by a harmonic
oscillator 
with a coordinate  $\eta$ and frequency $\Omega$. 
The internal oscillator 
is coupled to a  free 
scalar field $\phi$.
The total effective action is 
\beq
S = -M\int d\tau  -  qE_x\int X dt +
 {1\over2}\int\biggl(\Bigl({d\eta\over d\tau}\Bigr)^2 
-\Omega^2\eta^2\biggr) d\tau + \int g_0 \eta
\phi(X(t(\tau)),t(\tau))d\tau
+ S_F.
\eeq
Here, $\tau$ is the proper time in the detector's rest
frame, 
$X$ is the position of the detector,
$g_0$ is the coupling strength with 
a scalar field $\phi$ and $S_F$ is the action of the
field.
Since we would like to describe the back reaction on
the trajectory
let us rewrite this action in terms of the
 inertial frame time $t$. The action of the 
accelerated detector is then given by 
\beq
 \int \biggl[\Bigl(-M - g_0\eta\phi(X,t) 
\Bigr)\sqrt{1-\dot X^2} -
qE_x X\bigl] dt    + 
{1\over2}\int\biggl[{1\over\sqrt{1-\dot X^2} } 
\Bigl({d\eta\over dt}\Bigr)^2 - \sqrt{1-\dot X^2}
\Omega^2\eta^2 \biggr]dt.
\eeq
This yields a simple expression for the Hamiltonian of
the total 
system with respect to the inertial frame:
\beq
H = \sqrt{P^2 + M_{eff}^2}  -  qE_xX + H_F, 
\label{1}
\eeq
where the effective mass $M_{eff}$ is given by  
\beq
M_{eff}= M + {1\over 2} \Bigl( \pi_\eta^2 + \Omega \eta^2
\Bigr) 
+ g_0 \eta \phi(X),
\eeq
and $\pi_\eta = {\partial L \over \partial\dot \eta} =
\dot \eta/
 \sqrt{1-\dot X^2}$.
The validity of our model rest upon a the assumption
that the Schwinger
pair creation effect can be neglected for our detector.

Since the Schwinger pair creation process 
is damped by the factor  $\exp(-\pi M^2/qE_x)$ this 
implies the limitation $M^2>qE_x$. Notice that since 
the acceleration is $a=qE_x/M$, this implies that  
$M> a = 2\pi T_U$.
In the following we set $E_x=1$ for convenience.

To obtain a quantum mechanical model we simply need to 
impose quantization conditions  on the 
conjugate pairs $X, P$ and $\eta, \pi_{eta}$ and use the
standard 
quantization procedure for the scalar field. 
It is convenient to introduce internal energy level 
raising and lowering operators $A^\dagger$ and $A$. 
The harmonic oscillator Hamiltonian
can then be replaced by $\Omega A^\dagger A\equiv
\Omega N$ 
and the internal coordinate
by $\eta=i(A^\dagger-A)/\sqrt{2\Omega}$. 
This form can also be used in other, more general
cases, however the 
simple commutation relation  $[A, A^\dagger]=1$
in the case of a harmonic oscillator, needs to be
modified 
accordingly.  

So far we have not imposed a limitation on the 
coupling strength $g_0$. In the 
case of small coupling $g_0(t) = \epsilon(t)$ the 
Hamiltonian can be written to first order in
$\epsilon(t)$ as 
\beq
H = H_D - qX  + H_F + H_I.
\eeq
Here 
\beq
H_D = H_D(P, N) = \sqrt{P^2 + (M+\Omega A^\dagger A)
^2}
\eeq
is the free detector Hamiltonian, 
$H_F$ is the free field Hamiltonian
\beq
H_F= {1\over2}\int dx' [\Pi_{\phi}^2 +
(\nabla\phi)^2+m_f^2\phi^2 ],
\eeq
and 
\beq
H_{I} =
 i\epsilon(t)  \biggl\lbrace  {m_N\over H_D},  
(A^\dagger-A) \phi(X,t) \biggr\rbrace,
\eeq
where $m_N \equiv M +N\Omega= M +\Omega A^\dagger
A$ and
 the anti-commutator,
$\lbrace A,B\rbrace ={1\over2}(AB+BA)$, maintains
hermiticity.
We have also absorbed a factor of $1/\sqrt{2\Omega}$ in
the definition
of $\epsilon(t)$.
Comparing this interaction term with  
that used in the absence of a back-reaction we note
that apart from the  appearance of an  anti-commutator there
is also a new factor ${m_N\over H_D}$. As we shall see,
it corresponds to an operator boost factor  from the 
inertial rest frame to the detector's rest frame.

In the Hiesenberg representation   the eqs. of motion
for the 
detector's coordinates $X$ and $P$ are given by: 
\beq
\dot X = {P\over H_D}
 - i\epsilon(t)  \Bigl\lbrace {m_N P\over H_D^3},
(A^\dagger-A)\phi(X) \Bigr\rbrace, 
\label{X}
\eeq
\beq
\dot P = q  
-i\epsilon(t)  \biggl\lbrace  {m_N \over H_D},  
(A^\dagger-A) \phi'(X,t) \biggr\rbrace,
\label{P}
\eeq
where $\phi' = {\partial\phi\over\partial x}$.
We also have
\beq
\dot A = -i  \bigl(H_{D,N+1} - H_{D,N}\Bigr)A  -i[A,
H_I],
\label{A}
\eeq
and
\beq
(\Box - m_f^2)\phi(x,t) =
 i\epsilon(t) \biggl\lbrace  {m_N\over H_D}, 
 (A^\dagger-A) \delta(x-X) \biggr\rbrace .
\eeq

In the zeroth order approximation ($\epsilon=0$)
the solution of eqs. (\ref{X}-\ref{A}) is 
\beq
X^{(0)}(t)= X_0 +{1\over q} \Bigl[H_D(t)-
H_D(t_0)\Bigr], \ \ \ \
P^{(0)}(t) = P_0 +q(t-t_0),
\eeq
\beq
H_D^{(0)}(t) = \sqrt{ (P_0+q(t-t_0))^2 +(M+\Omega 
A^\dagger_0 A_0)^2},
\eeq
and 
\beq
A^{(0)}(t) =  \exp \Bigl[
-i\int_{t_0}^t ( H_{D,N_0+1}- H_{D,N_0})dt'  \Bigr]A_0.
\label{At}
\eeq
Here the subscript was used to denote the operator at 
time $t=t_0$ and the superscript to denote 
the zeroth order solution. To simplify notation we
shall drop  
the superscript.
Notice that $N_0 = A_0^\dagger A_0$ is a constant of
motion in the zeroth
order approximation.

It is now useful to introduce a proper time $operator$
$\tau(t)$:
\beq
\tau(t)=   {M+\Omega A^\dagger A \over q}
\sinh^{-1}\Bigl[{q(t-t_0)+ P_0\over M+\Omega
A^\dagger A}\Bigr].
\label{taut}
\eeq
The factor  $({M+\Omega A^\dagger
A)/H_D}= {m_N/ H_D}= {d\tau\over dt}$ 
appearing in the coupling to the field (eq. 8) can therefore be
interpreted as an operator boost factor 
${d\hat \tau(t)\over dt}$, from the inertial frame to
the detector's rest frame. Notice that $\tau$ depends only
on $P_0$ and $N$.

In terms of the proper time operator, the detector's
trajectory can
be simplified to:
\beq
t- t_0 - \tilde T_0 = {1\over a}\sinh a\tau,
\label{tt}
\eeq
\beq
X-\tilde X_0 = {1\over a} \cosh a\tau,
\label{xt}
\eeq
where 
\beq
\tilde T_0 = -{P_0 \over q} \ \ \ \ \ \ \ \tilde X_0 =
-{H_D-qX\over q},
\eeq
and  the acceleration $a$ is given by the $operator$
\beq
a = a_N= {q\over M + \Omega A^\dagger A} = {q\over
m_N}.
\eeq
The operators $\tilde T_0$ and $\tilde X_0$  determine 
the location  of the Rindler coordinate system of the
detector with 
respect to 
the Minkowski coordinates ($t,x$). The space-time
location of the intersection point
of the future and past Rindler horizons is given by 
$(-t_0 -\tilde T_0, -\tilde X_0)$. Since 
\beq
[\tilde X_0, \tilde T_0 ] = {i\over q},
\label{TXtilde}
\eeq
the location of this  space-time point becomes quantum
mechanically
smeared.
 
Another set of useful operators \cite{qframe} we shall
introduce is that 
of the location
of the future and past Rindler horizons $\h_+$ and
$\h_-$, respectively.
They can be found from the relations 
\beq
\h_+(t)=\lim_{t\to \infty}X(t), \ \ \ \ \ \ \ \ 
\h_-(t) =\lim_{t\to-\infty}X(t).
\eeq
We find 
\beq
\h_+(t) = -\tilde T_0 +\tilde X_0 +(t-t_0) = {P(t)\over
q} - {H_D-qX\over q},
\label{hp}
\eeq
and 
\beq
\h_-(t) = \tilde T_0 +\tilde X_0 -(t-t_0) = -{P(t)\over
q}- {H_D-qX\over q},
\label{hm}
\eeq
Therefore we can express $X(t)$ as  
\beq
X(t) = \h_+(t) + {1\over a} e^{-a\tau} \ \
 \stackrel{{t\to\infty}}{\to} \ \ \h_+(t),
\eeq
and
\beq
X(t) = \h_-(t) +  {1\over a} e^{a\tau} \ \
\stackrel{{t\to-\infty}}{\to} \ \ \h_-(t).
\eeq
In terms of $\h_\pm$, the Hamiltonian of the detector
in 
an external electric field has the simple form:
\beq
H_D - qX = -{q\over 2} \Bigl( \h_+ + \h_-
\Bigr).
\eeq
Finally, $\h_\pm$ satisfy the commutation  relation:
\beq
[\h_-, \h_+] = 2{i\over q}
\label{hpm}
\eeq
Examining eqs. (\ref{TXtilde}) and (\ref{hpm}), we 
notice that since $q =  a M$,
in the limit of  constant 
acceleration but large mass, the commutators   
vanish as $M^{-1}$ and the classical 
trajectory limit is restored.

\section{The Transition Amplitude}

We shall now proceed to calculate the first order 
transition amplitude between the internal energy levels 
$n$ 
and  $n+1$ of the detector.
To this end it will be most convenient to  use 
the interaction representation.
The operators in this representation are the solutions
of the 
free equations of motion given by
(\ref{At},\ref{taut},\ref{tt},\ref{xt}),
and the wave function satisfies   the Schr\"odinger
equation
\beq
i\partial_t |\Psi\ra = H_I |\Psi\ra.
\eeq
Given at $t=t_0$ by the initial wave function 
$|\Psi_0\ra$, to first order in $\epsilon$  
the final state at time $t$ is given by 
\beq
|\Psi(t)\ra = \biggl[1-i\int_{t_0}^t \
\epsilon(t')  \biggl\lbrace  {m+\Omega
A^{\dagger}A\over H_D},  
i(A^{\dagger}-A) \phi(X,t') \biggr\rbrace dt'  \biggr]
|\Psi(t_0)\ra. 
\eeq

Let us set initial conditions for the internal
oscillator 
to be in the $n$'th exited 
state $|n\ra$,
and for the scalar field to be in a 
Minkowski vacuum state $|0_M\ra $.
The initial
state of the total system is therefore given by 
$|\Psi(t_0)\ra = |0_M\ra\otimes|n\ra\otimes|\psi_D\ra$,
where $|\psi_D\ra$
denotes the space component of the  detector's wave
function.
Using the solution (\ref{At}) for $A$ and $A^\dagger$,
the transition amplitude can be expressed as:
$$
|\Psi(t)\ra = |\Psi(t_0)\ra 
-{\epsilon\over2}\int_{t_0}^t dt'\biggl[
$$
$$
 \sqrt{n+1}|n+1\ra
\biggl( {m_{n+1}\over H_{D,n+1}}
e^{i\int_{t_0}^{t'}\Delta H_{n+1}dt''}
\phi(X_n(t'),t')+
e^{i\int_{t_0}^{t'}\Delta H_{n+1}dt''}
\phi(X_n(t'),t'){m_n\over H_{D,n}} \biggr)
$$
$$
-\sqrt{n}|n-1\ra 
\biggl( {m_{n-1}\over  H_{D,n-1}}
e^{-i\int_{t_0}^{t'}\Delta H_{n}dt''}\phi(X_n(t'),t')+
e^{-i\int_{t_0}^{t'}\Delta H_{n}dt''}\phi(X_n(t'),t'){m_n
\over H_{D,n}} \biggr)
$$
\beq
\biggr]\otimes|0_M\ra\otimes|\psi_D\ra.
\eeq
Here we used the notation
$\Delta H_{n} = H_{D,n}- H_{D,n-1}$.
 The subscript $n$ (e.g. in $X_n(t')$),  means that we need
to substitute 
the free solutions with  $N=n$.
In two dimensions the solutions for a free massless
scalar field 
can always be separated into right and left moving
waves, i.e.
$\phi = \phi_L(V)+ \phi_R(U)$ where $U=t-x, \ \ V=t+X$.
For simplicity we will limit the discussion to massless
scalar fields and examine 
the solution  only for right moving waves.
Therefore, we substitute for $\phi$: 
\beq
\phi_R(U) = \int_0^\infty {d\omega\over \sqrt{4\pi\omega}}
\Bigl(e^{-i\omega U}
a_\omega + e^{i\omega U}a^\dagger_\omega \Bigr).
\eeq
Using eqs. (\ref{tt},\ref{xt},\ref{hp}) we find that on
the trajectory of the detector the light cone
coordinate $U$ is given by
\beq
U|_{D} = t- X = -\h_{+0}-t_0 -{1\over a} e^{-a\tau}.
\eeq

Neglecting the  constant phase factor $\exp(i\omega t_0)$, the 
final state can be written as:
$$
|\Psi(t)\ra = |\Psi(t_0)\ra  
-{\epsilon\over2}\int{d\omega\over\sqrt{4\pi\omega}}
 \int_{t_0}^t dt'\biggl[ 
$$
$$\sqrt{n+1}|n+1\ra
\biggl( {m_{n+1}\over H_{D,n+1}}e^{i\int\Delta
H_{n+1}dt''}
e^{i\omega(-\h_{+0n} -{1\over a_n} e^{-a_n\tau_n})}
+
e^{i\int\Delta H_{n+1}dt''}
e^{i\omega(-\h_{+0n} -{1\over a_n} e^{-a_n\tau_n})}
{m_n\over H_{D,n}} \biggr)
$$
$$
-\sqrt{n}|n-1\ra 
\biggl( {m_{n-1}\over H_{D,n-1}}e^{-i\int\Delta
H_{n}dt''}
e^{i\omega(-\h_{+0n} -{1\over a_n} e^{-a_n\tau_n})}
+
e^{-i\int\Delta H_{n}dt''}
e^{i\omega(-\h_{+0n} -{1\over a_n} e^{-a_n\tau_n})}
{m_n \over H_{D,n}} \biggr)
$$
\beq
\biggr]\otimes|1_{\omega M}\ra\otimes|\psi_D\ra.
\label{tamp}
\eeq

This is an exact result in the first order
approximation
in $\epsilon$. So far we have not introduced 
additional assumptions on $M$, $\Omega$ or
$a_n=q/m_n$.
We shall now apply a large mass limit.  We shall assume
that 
\beq
M  >> a_0 = {q\over M}.
\eeq
This restriction is indeed equivalent to a suppression of the 
Schwinger pair production process. 
Since for the Unruh radiation we need only detector energy gaps
with $\Omega= O(a/2\pi)$,  we also require 
\beq
M >> \Omega, 
\label{small-omega}
\eeq

Under these assumptions let us proceed to simplify expression
(\ref{tamp}).

First consider the term $\exp(\int \Delta H_{n+1}dt)$:
$$
i\int_{t_0}^t \Delta H_{n+1}dt' = 
i\Omega\int_{t_0}^t {m_n\over H_{D,n}}
\Bigl[
1+{1\over2} {\Omega\over m_n}{P^2\over H_{D,n}^2}\Bigr]
+ O(\Omega^3/M^3)
$$
$$
= {i\Omega\tau_n} + {i\over2}\Omega^2
\Bigl[{1\over m_n}\tau_n - {1\over q}\tanh(a_n\tau_n)
\Bigr]
+ c(P_0)+ O(\Omega^3/M^3)
$$
\beq
\simeq
 i\Omega\Bigl(\tau_n+ {1\over2} {\Omega\over m_n}\tau_n
- {\Omega\over 2q}(1-2\exp(-2a_n\tau_n)) \Bigr) +c(P_0)+
O(\Omega^3/M^3),
\eeq
where $c(P_0)$ is a constant, and  in the last line  we
have used the large $\tau$ approximation. 
This approximation is justified since the transition
amplitude is
dominated by contributions arising from integration
over
large $\tau$.
In the following we shall hence neglect the exponential
correction
and the constant $c(P_0)$ which gives rise only to an
overall 
phase, and use the approximation:
 \beq
i\int_{t_0}^t \Delta H_{n+1}dt' = 
i\Omega\tau_n\Bigl(1+ {1\over2} {\Omega\over m_n}
\Bigr).
\label{intdh}
\eeq

Next consider the exponential terms in (\ref{tamp}) 
which contain the horizon operator  $\h_{+0}$. 
Only these terms maintain a dependence on 
the operator $X$ as  $\h_{+0} = X + G(P_0)$, where $G$
is a function of $P_0$ only. Therefore, $[\h_+, P_0] = i$.
Noting that the second term in the exponential depends only on
$P_0$,  and using  the Campbell-Baker-Hansdorff identity 
yields:
\beq
\exp[{-i\omega(\h_{+0} +{1\over a_n} e^{-a_n\tau_n})}]= 
\exp\Bigl[
{{i\over2 q}\omega^2 e^{-a_n\tau_n}{m_n\over H_{D,n}}
+O(M^{-2}a_0^{-2})}
\Bigr]
\exp({-i\omega{1\over a_n} e^{-a_n\tau_n}}) 
\exp({-i\omega\h_{+0}})
\label{exph}
\eeq

Notice that  the unitary
operator $e^{-i\omega\h_{+0}}$ generates  the
translation:
$p_0 \to p_0 +\omega$. This corresponds to a recoil of the
detector and ensures  total
momentum conservation when the detector is exited and emits 
a Minkowski photon.

Finally, we consider the boost operator: 
\beq
{m_{n+1}\over H_{D,n+1}} = {m_{n}\over H_{D,n}}
\biggl[1 +{\Omega\over m_{n}}
\Bigl(1-{m_n^2\over H_{D,n}^2}\Bigr) + O(\Omega^2/M^2)
\biggr].
\eeq
Since for large $\tau$ 
\beq
{m_n\over H_{D,n}}= {1\over \cosh(a_n\tau_n)} =
 2 e^{-a_n\tau_n} - O( 2e^{-3a_n\tau_n}),
\eeq
we  shall approximate this boost factor by
\beq
{m_{n+1}\over H_{D,n+1}} =  {m_{n}\over H_{D,n}}
\biggl[1 +{\Omega\over m_{n}} \biggr].
\label{bf}
\eeq

We can now return to the transition amplitude
(\ref{tamp}) 
and for simplicity focus only on the  amplitude 
$A(\omega,n+1,p)=
\la 1_\omega, n+1, p|\Psi(t)\ra$. The indices $\omega, n+1,p$
correspond to the outgoing states of the photon, internal
detector levels and to the detector's momentum, respectively.  
Using eqs.  (\ref{intdh},\ref{exph},\ref{bf}) we find 
\beq
A(\omega,n+1,p)=
 -{i\epsilon\over2}\sqrt{n+1\over 4\pi\omega}
\int^t_{t_0} dt' 
\biggl[
\biggl({m_{n}\over H_{D,n} (p_0+\omega)}
+{m_{n}\over H_{D,n}(p_0)}\Bigl(1+{\Omega\over
m_n}\Bigr)\biggr)\times
\label{int}
\eeq
$$
\exp\bigg(i\Omega\Bigl(1+{1\over2}{\Omega\over
m_n}\Bigr) 
\tau_n -         {i\omega{1\over a_n} e^{-a_n\tau_n}}  
+{{i\over2 q}\omega^2 e^{-a\tau_n}{m_n\over H_{D,n}} }
\biggr)\biggr]\phi_D(p+\omega).
$$
Here, $\phi_D(p)= \la p|\psi_D\ra$. 
To obtain (\ref{int}) we  used a representation with 
$\h_{+0}$ and $P_0$ 
as conjugate operators, and used  the unitary
operator
$\exp-i\omega\h_{+0}$ to generate translations in the
momentum. 
At this point the
transition amplitude is expressed as a c-number
integral. 

Let us proceed to investigate this integral.
For large $t$ 
the phase $\theta$ of the integrand can be approximated
by
\beq
\theta= \Omega\Bigl(1+{1\over2}{\Omega\over m_n}\Bigr)
\tau_n -  \omega{1\over a_n} e^{-a_n\tau_n}  
+{1\over q}\omega^2 e^{-2a_n\tau_n}.
\label{phase}
\eeq
The stationary phase condition yield 
\beq
\omega = - \Omega\biggl(1- {\Omega\over m_n} +
 O (\Omega^2/M^2) \biggr)
\  e^{a_n\tau_n}.
\label{sphase}
\eeq
This can be compared with the case of a classical
trajectory obtained by sending $m\to \infty$.
In the present case, the  frequency at the stationary
point is shifted by ${\Omega^2\over m_n} e^{a_n\tau_n}$, which is
in agreement with ref. \cite{parentani} up to a numerical factor
of $1/2$.
As long as  ${\Omega\over m_n}<1$, the correction is small and
the saddle point frequency remains exponentially high.

A second phase appears in the amplitude due to the shift in the 
momentum of the particle. Let the initial wave function of the 
detector be in an eigenstate of momentum, $|\psi_D\ra= |k\ra$.
The horizon shift operator acting on the state yields 
$$
e^{-i\omega\h_{+0} } |k\ra = e^{-i \omega X_0} 
e^{i\omega  T_0} e^{-i\omega^2/2q} |k\ra 
$$
$$
= e^{-i(\omega k + \omega^2/2)/q} e^{ -i\omega \tilde X_0} |k\ra
$$
\beq
= e^{-i(\omega k + \omega^2/2)/q}  |k - \omega \ra
\label{recphase}
\eeq
where in the first line we have used the Campbell-Baker-Hausdorff
identity. 
The phase, $(\omega k +\omega^2/2)/q$,  is identical to that obtained
by Parentani \cite{parentani}, (the factor $\omega^2/2$ arises
here from the non-commutativity of $X_0$ and $T_0$) . 
Here, the phase 
is directly obtained from the shift 
generated by the future horizon operator
as a consequence of the recoil due to an emission of a scalar
photon. This recoil is further discussed in Section IV.

Next consider the recoil affects on the boost factor 
 ${m_{n}\over H_{D,n}(p+\omega)}$ in eq. 
(\ref{int}).
The shift of  $p\to p +\omega$ in this  boost factor,
is equivalent to a shift in time given by 
$t\to t' = t+{\omega\over q}$. In terms of the proper
time (which is now a c-number) this 
correspond to the transformation
\beq
\tau_n \ \ \to \ \ \tau'_n = \tau_n + {\omega\over
q}e^{-a_n\tau_n}
\eeq
For  transitions with $\tau_n(t)-\tau_n(t_0)>>1/a_n$, 
this transformation does not
modify the integral. Hence in terms of $\tau'_n$:
\beq
{m_{n}\over H_{D,n}(p+\omega)}= {d\tau'_n\over dt}.
\eeq

The second, unshifted,  boost factor can be expressed
in terms of $\tau'_n$ as 
\beq
{m_n\over H_{D,n}}\Bigl(1+ {\Omega\over m_n}\Bigr)=
{d\tau'_n\over dt}\Bigl( 1 
+{1\over m_n}(\Omega + \omega e^{-a_n\tau'_n})
+O(\Omega^2/M^2) \Bigr).
\eeq
Hence by expressing the 
integral (\ref{int}) in terms of  $\tau'_n$ we find that
the two terms are equal up to order $O(\Omega^2/M^2)$ and
an additional piece  that (up to this order) vanishes
at the stationary point (\ref{sphase}).

Expressing the phase in terms of $\tau'$ we find 
\beq
\theta=\Omega\Bigl(1+{1\over2}{\Omega\over
m_n}\Bigr)\tau_n'
 -{\omega\over a_n} \Bigl(1+{\Omega\over m_n}\Bigr)
e^{-a_n\tau_n'} +O(\Omega^2/M^2). 
\eeq
where the term involving ${\omega^2\over q}e^{-2a_n\tau_n}$
in eq. (\ref{phase}) has dropped out and we 
are left only with the higher order corrections
$O(\Omega^2/M^2)$, which will be neglected. 

In terms of $\tau'_n$ the amplitude $A(\omega,n+1,p)$
can be written as:
\beq
 -{i\epsilon}\sqrt{n+1\over
4\pi\omega}\phi_D(p+\omega)\Biggl[
\int
d\tau_n'\exp\bigg({i\Omega\Bigl(1+{1\over2}{\Omega\over
m_n}\Bigr)\tau_n'} -i\omega{1\over
a_n}\Bigl(1+{\Omega\over m_n}
\Bigr) e^{-a_n\tau_n'}  \biggr) + {\xi\over m_n} \Biggr]
\eeq
where 
\beq
\xi ={1\over 2} \int d\tau_n(\Omega +\omega
e^{-a_n\tau_n})\exp\Bigl(
i\Omega\tau_n - i{\omega\over a_n}(1+ {\Omega\over
m_n})e^{-a_n\tau_n} \Bigr)
\eeq
For large $\tau_n $,  $\xi\sim O({\Omega\over M})$, and   
the term $\xi/m_n$ can be
neglected.

Finally we obtain: 
\beq
A(\omega,n+1,p) = {i\epsilon}\sqrt{n+1\over
4\pi\omega}\phi_D(p+\omega)  a_n^{-1}
\Bigl(  {\omega\over a_n}(1+ {\Omega\over
m_n})\Bigr)^{i{\Omega'\over a_n}}
\Gamma(-i{\Omega'\over a_n}) e^{-{\pi\Omega'/2a_n}} 
+O(\Omega^2/M^2),
\label{amplitude}
\eeq
where $\Gamma$ is the Gamma function, and 
\beq
\Omega' = \Omega\Bigl( 1+ {1\over2} {\Omega\over m_n}
\Bigr)=\Omega\Bigl( 1+ {1\over2} {\Omega\over M}  \Bigr)
+O(\Omega^2/M^2) .
\eeq

This transition amplitude seems similar to the Unruh amplitude 
obtained in the absence of recoil and quantum smearing. 
Our amplitude  agrees, (when $\Omega/M$ corrections are neglected),
with the
result obtained by Parentani \cite{parentani}. 
 As we have already seen, the momentum dependent 
phase relevant to the
decoherence process \cite{parentani}, is also precisely  obtained
in our method (eq. \ref{recphase}).

\section{Recoil and Quantum Smearing}

The  purpose of this section is to give a qualitative simple
picture of the recoil. We shall show how this process can 
be simply  expressed in terms of the effect of horizon shift
operators on the detector's wave function.
Since as we shall see, the recoil involves  exponentially large
shifts, in this section we can neglect the $1/M$ corrections.

Let us re-state the results of the last section in a
more qualitative
way. For the case of a classical trajectory, it was
shown by  Unruh and Wald \cite{uw} that if the detector is
initially in  the 
ground state then the 
final state can be written as 
\beq
|\Psi(t)\ra = |\Psi(0)\ra -i|n=1\ra\otimes
a_{R\Omega} |0_M\ra.
\eeq
Here, $a_{R\Omega}$ is the annihilation 
operator of a quantum with frequency $\Omega$
with respect to the Rindler coordinate 
system that is defined by the detector's trajectory.
Using the well known relation \cite{unruh} of
$a_{R\Omega}$  
to Minkowski  creation and annihilation operators 
$a_{M}$ and $ a_{M}^\dagger$, 
they get  
\beq
|\Psi(t)\ra = |\Psi(0)\ra -i C(\Omega,a)|n=1\ra
{e^{-\pi\Omega/2a} \over 
(e^{\pi\Omega/a} -e^{-\pi
\Omega/a})^{1/2}} a_{M}^\dagger |0_M\ra,
\label{uw}
\eeq
where  $C$ is a normalization factor. 
Note that $a^\dagger_{M}$ creates a positive frequency 
Minkowskian
photon,  which is not in a state of definite frequency
$\omega$.
Qualitatively we can use the stationary phase
approximation 
eq. (\ref{phase}) to relate the typical frequency 
of this photon to the time of emission $\tau$.

We can now  use the result obtained  in the last
section
to replace eq. (\ref{uw}) with 
$$
|\Psi(t)\ra = |n=0,\psi_D,0_M\ra 
$$
\beq
-i C(\Omega,a_n)|n=1\ra
{e^{-\pi\Omega/2a} \over 
(e^{\pi\Omega/a} -e^{-\pi
\Omega/a})^{1/2}}\Bigl( 
e^{-iH_F \h_{+}}a_{M R}^\dagger + e^{+iH_F \h_{-}}
a_{M L}^\dagger
\Bigr)
|0_M,\psi_D\ra
\label{shift}
\eeq  

Here we have restored the full coupling with the left and right
moving waves. 
The operators $a_{M R}^\dagger $ and $a_{ML}^\dagger$,
 correspond to creation operators of  right and left
moving waves  respectively.
This equation can be easily generalized to the case of
transitions between 
any two levels $n$ to $n+1$, as well as to the case of
de-excitations.
We have assumed that the scalar field is massless.
However, for 
a massive field
we simply need to replace $e^{-iH_f \h_{+0}}$ by
$e^{iH_f\tilde T_0 -iP_f 
\tilde X_0}$ etc.

The new feature of eq. (\ref{shift}) is  
the insertion of the horizon shift operators $\exp( \pm
i H_F \h_\pm)$
which act on the wave function of the detector and of
the scalar field.
These shift operators generate correlations between
the ``emitted'' Minkowski scalar photon and the
trajectory of the 
detector.

To illustrate  these correlations, let us concentrate
only on the 
left moving  waves and express $a_{MR}^\dagger$ in
terms of creation
operators of definite  Minkowski frequency:
\beq
a_{MR}^\dagger = \int f(\omega) a_\omega^\dagger
d\omega
\eeq  

Eq. (\ref{shift}) can now be written as 
\beq
|\delta\Psi\ra = -iC'|n=1\ra\int d\omega dh_{+}
e^{-i\omega h_{+}}
f(\omega)\psi(h_{+}) |1_\omega\ra\otimes|h_{+}\ra.
\label{ent}
\eeq
Here we used a basis of  
$\h_{0+}$: $\h_+|h_+\ra = h_+|h_+\ra$. 
We see that the recoil interaction generates 
correlations between the shift 
$h_+$ in the $u$-time of the right moving ``emitted''
Minkowski photons with the ``horizons states''
$|h_+\ra$ of 
the detector. Therefore, the 
effect of ``horizon smearing'' yields, after
emission, the  
 final entangled state (\ref{ent}).
In each component of this state, 
the Unruh effect is manifested, with the correction
discussed in the previous section.
Since the corrections  do not depend on the uncertainty
or the smearing $\Delta h_+$ of the future event
horizon, the overall 
wave function still manifests the Unruh effect.

In order to examine the effect of the emission on the
detector we 
can re-write eq. (\ref{ent}) by using as a basis the
past horizon
operator $\h_-$. We obtain:
\beq
|\delta\Psi\ra= -iC'|n+1\ra\int d\omega dh_- 
f(\omega)\psi(h_-) |1_\omega\ra\otimes|h_--\omega\ra,
\eeq
where $\psi(h_-) = \la h_-|\Psi_D\ra$.
Since $\h_\pm$ are conjugate operators, the  operator 
$\exp-i\omega\h_+$  has shifted the past horizon
operator by $\omega$.
It is interesting  to notice that the shift by $\omega$
of the past horizon can be exponentially large.
In fact, from the stationary phase approximation we get
that 
it is related to the time of emission $\tau$ as:
$\Omega\simeq \Omega\exp (a\tau)$.
Therefore a  detection of a particle of energy $\Omega$
generates an exponential shift in the location of the
past horizon
of the detector: 
\beq
\delta h_- = h_{-out} -h_{-in}  \simeq \Omega \exp
(a\tau)
\label{dhm}
\eeq
The meaning of this shift is as follows.
We can use the initial state $\psi_{in}$ to define the
location 
$h_{-in}$ of the past horizon.
We can also use the final state $\psi_f$ of the
detector 
and by propagating it
to the past (with the free Hamiltonian) 
determine the location $h_{-out}$.
These two locations differ by an exponential shift.
    
The propagation of a wave function to the past
might seem strange. However the same phenomenon occurs if the 
detector is excited in the past at $\tau<0$.
In this case it emits a left moving Minkowski particle.
We find that this induces an 
exponentially large shift in the location of the future
event 
horizon operator
$\h_+$:
\beq
\delta h_+ = h_{+out} - h_{-in} \simeq \Omega \exp
(-a\tau)
\label{dhp}
\eeq

The manifestation of the back reaction as an
exponentially large
shift is related to the method  of 
't Hooft \cite{thooft} and of  Schoutens, Verlinde and  
Verlinde
\cite{verlinde}. In their case,  
infalling matter into the black hole, 
induces an exponential shift of  
the time of emission of the Hawking photon in the
future. 
The reason is that the Hawking photons stick so close
to the horizon that even a small shift of the horizon 
still modifies the time of emission.
In our case this exponential shift is related to the 
exponential energy of the emitted Minkowski photon.
In both cases, the back reaction requires the existence
of exponentially
high frequencies in the vacuum.
As in the case of Hawking radiation, a naive cutoff  
eliminates the thermal spectrum seen by the Unruh
detector.

\vspace {3cm}

{\bf Acknowledgment}

I wish to thank W. G. Unruh for  discussions and very
helpful comments.
 I have also benefited from discussions with 
  S. Nussinov and J. Oppenheim.

\vfill \eject


\begin{thebibliography}{99}



\bibitem{hawking}
S. Hawking , Nature {\bf248}, 30 (1974).
S. Hawking, Commun.  Math. Phys. {\bf43}, 199 (1975). 


\bibitem{unruh}
W. G. Unruh, Phys. Rev. D {\bf14}, 870 (1976).

\bibitem{rev}
For a recent review see:
R. Brout, S. Massar, R. Parentani, and Ph. Spindel, 
Phys. Rep. {\bf260}, 329 (1995).

\bibitem{parentani}
R. Parentani, Nucl. Phys. B {\bf 454}, 227 (1995). gr-qc/9502030.
also see: R. Parentani, S. Massar, Phys.Rev.D55, 3603,1997. hep-th/9603057. 

\bibitem{qframe}
S. Popescu and B. Reznik,  Unpublished.

\bibitem{uw}
W. G. Unruh and R. M. Wald, Phys. Rev. D {\bf29}, 1047 
(1984).

\bibitem{thooft}
G. 't Hooft, Nucl. Phys. B {\bf 256}, 727 (1985)
\bibitem{verlinde}
K. Schoutens, E. Verlinde and H. Verlinde,
Phys. Rev. D {\bf48}, 2670 (1993).

\end{thebibliography}
\end{document}